# Gaming Together on Discord: Teen Gamer’s Cross-Platform Practices

Elena Koung*

The Pennsylvania State University, elenakoung@psu.edu

Xinning Gui

The Pennsylvania State University, xinninggui@psu.edu

Yubo Kou

The Pennsylvania State University, yubokou@psu.edu

Discord is one of the most popular communication platforms among gamers. While prior research has highlighted its role in community building, relatively little attention has been paid to its original gaming context—how it shapes gameplay and social experiences. To address this gap, we conducted semi-structured interviews with 16 teenage Discord users. Through reflexive thematic analysis, we show how players leverage Discord to create more collaborative and socially enriched experiences that extend beyond the game itself. However, gaming together on Discord also resulted in social and security risks. We conceptualize gaming together on Discord as a cross-platform practice that extends gameplay beyond a game and supports players’ social needs. Additionally, cross-platform practice also introduces the ‘platform gap,’ where fragmented governance between platforms exposed players to risks. To address this tension, we propose design implications aimed at bridging the platform gap, strengthening communication channels, and supporting safer gaming experiences.

## 1 INTRODUCTION

With over 21 million gaming-related servers, Discord is a social platform for gamers to find online communities that support their diverse gaming interests [7]. Discord is often used alongside gaming, where players can seamlessly access the same Discord communities across devices, enabling shared communication regardless of hardware [66]. While conversations take place on Discord, gameplay is bounded within individual gaming platforms. HCI and game research has examined player experiences in specific games (e.g., Roblox [74, 75]) or genres such as MOBAs [65, 98, 149] and first-person shooters [30, 134]). More broadly, playing multiple games and using different hardware means that player behavior is inherently distributed across platforms [117, 118]. Discord brings these distributed gaming experiences together [155], functioning as a hub where gamers gather, communicate, and sustain play beyond individual game boundaries.

Discord is popular among teenage gamers [63], providing them with a casual space to explore their gaming interests, socialize through memes, and talk about their daily lives [70]. Its communication features have been shown to support collaboration, learning and community-building across contexts [10, 32, 79, 101, 132]. Notably, Discord has been used as a site for engaging teens in HCI research [103] and as a space to practice restorative justice [41]. Thus, Discord's communication affordances enable teens to participate in online gaming communities.

However, teenagers must navigate various risky online social situations within Discord [76] and interact across multiple platforms [45, 133, 145], yet we know little about how their practices across gaming platforms reshape existing safety dynamics or introduce new vulnerabilities. Although the risks linked to teenage gaming, including addiction [84, 146], gambling [56], health [38], and privacy [26], are well documented, far less attention has been paid to risks that emerge across platforms when teens combine gameplay with social interaction on platforms such as Discord. Thus, understanding these dynamics is critical for identifying risks that arise from the concurrent usage of multiple platforms. We conducted 16 interviews with teen gamers who use Discord to answer the research question:

**RQ: How does Discord impact teenagers' social experiences when gaming?**

Our research addresses how Discord is used while gaming, particularly in how teenagers socially interact with others on multiple platforms. We contribute to CHI Play and HCI research by broadening the examination of players' social experiences beyond a single game title, highlighting cross-platform participation within gaming communities. We show how Discord extends teenagers' gaming participation beyond in-game play, enabling continuous social interaction and coordinating gameplay across platforms. Further, the movement between platforms highlights a "platform gap," where no single moderation system has full visibility or control, allowing abusive behavior to persist across environments.

## 2 BACKGROUND

Discord shapes how players communicate because gaming is becoming more social [109]. It is a communication platform that supports text, voice, and video communication organized through servers and channels, allowing players socialize when gaming without being face-to-face [44]. Rather than replacing in-game chat, Discord supplements it by providing a space that spans devices and gaming sessions.

Gamers inherently make up a substantial population of Discord users. The majority of Discord servers focus specifically on gaming and over 90% of Discord users report playing games while using the platform [61]. Moreover, Discord only requires players to be 13 or older to interact in shared gaming spaces, such as servers and voice channels [39]. Teenagers gravitate toward Discord because it offers both autonomy and social connectivity within gaming communities. The structure of Discord supports real-time communication, private messaging interactions, and broader community engagement, making it especially appealing to youth who view gaming as a primary site of social life beyond entertainment. Gaming sessions are prolonged significantly when players engage with friends and viewing a peer's livestream frequently motivates players to start the same game shortly afterward [61]. For example, players may join a voice call to watch a friend screenshare live gameplay before deciding to log on, remain connected through voice chat while playing, and continue discussing strategies or replaying highlights together after the session ends.

Discord's significance in gaming lies in its ability to bridge social interaction and gameplay across different titles and ecosystems. Discord's gaming communities illustrate the diversity of players' gaming preferences: the largest gaming server is *Marvel Rivals* is a hero shooter game with 4.1 million members, next is *Genshin Impact,* an RPG with 2.1 million members, and then *Blox Fruits* (a *Roblox* Game) with 1.75 million members [7]. While games provide interactions through gameplay within their own platform, Discord supports sustained community-building that is not confined to a single gaming platform.

## 3 RELATED WORK

### 3.1 Social Affordances from Third Party Platforms for Gaming

Social affordances in gaming are the game features and mechanics that enable, shape, or limit how players interact socially with one another [18, 33]. Social affordances foster social play [59], allowing multiple players to collaborate [37] and also interact in physical spaces for gaming activities together [68, 120, 142]. Besides social affordances within games, third-party platforms also offer social affordances outside of games [113]. Gaming experiences can be improved through third-party platforms, like Teamspeak and Discord, offering voice chat during gaming [123, 156]. For instance, Teamspeak and *Minecraft* could be used together for prosocial outcomes, where youth engaged in *Minecraft* were also developing leadership skills when chatting with others on Teamspeak [60, 127]. Although this research showed how third-party platforms may support certain positive social experiences, it is limited by the specific research context that overspecifies the game platform context for research observation.

Discord is a particular platform of interest for gaming and social interaction, as research often considers Discord to be a popular social media platform for gamers [24, 34, 87]. However, Discord affordances support more than social media. Families use Discord's social affordances, when gaming together at home and need to communicate via voice and audio [99]. Discord also supports fast-paced gameplay, improving responsiveness and teamwork through real-time voice coordination during compared to slower in-game text chat [17]. These affordances enables players to immediately coordinate, communicate, and organize on Discord, sustaining interaction even when they are not actively playing in a game environment [2]. Synchronous voice and video channels alongside asynchronous text communication allow members to sustain connection after leaving immersive environments [2, 71]. Notably, Discord supports friendships outside of the game context by facilitating event planning, ongoing conversation, and broader social engagement [92, 128]. In professional contexts, some Discord users may even manage social affordances to implement strategic boundaries that separate work from their gaming identity [94].

Discord supports gamers by providing a space for online community interaction. Structurally, Discord servers organize community-level communication through designated channels, typically organized by conversation topics [2]. Servers incorporate text and voice channels to facilitate discussion across a broad spectrum of subjects [71]. While some Discord users focus on talking about gameplay, others participate in humorous off-topic discussions and learning community values [128, 144]. Discord has also been used to increase community participation and build a space for Twitch streamers to interact with fans, strengthening viewers' attachment to the streamer [137]. Moreover, members of a server may participate in collective sensemaking, which allows members in a Discord server to handle misinformation collaboratively [144]. Additionally, members can step into community leadership roles, acting as moderators for a large community [40, 64]. Moderators from streaming and social VR platforms rely on Discord to satisfy their moderation needs, such as dealing with imminent threats [112, 113].

Across these studies, gamers used social affordances on Discord to play games and connect with others. Specifically, gamers may face challenges when using communication tools while gaming. Researchers found that audio-based communities had issues with integrating newcomers, managing participation, and resolving interpersonal conflict [113]. However, the focus on how third-party platforms, like Discord, naturally affects socialization and gameplay together remains less understood.

### 3.2 The Gaming Multi-Platform Ecosystem

The gaming multi-platform ecosystem describes the environment in which players use multiple platforms to fulfill their desired gaming experience. Platform studies allows for a focused analysis of the specific affordances and constraints that users interact with to enable or restrict expression [105]. Unlike the infrastructural lens, which focuses on ubiquitous media and sociotechnical systems [131], platform studies examines how the technical design of gaming ecosystems directly shapes user agency and interactions [105]. Further, platform studies highlights tensions between corporate interests and player motivations [83]. Thus, platform studies provides an appropriate lens for examining Discord and games together.

Multi-platform ecosystems allow users to leverage particular platforms for their specific affordances. Based on the games' specific affordances, players' gaming practices of "game-swinging" captures how multiple games are played in a single gaming session, where streamers on Twitch strategically switched between games they played [29]. Other practices in multi-platform ecosystems involve using forums to connect with an online community and participate in ongoing discussion about gameplay. [62, 110]. This provided personal benefits, such as increased gratification from self-presentation and relationship management [124].

Cross-platform play is an essential tool for community building because it allows players to connect regardless of hardware differences [116]. Research on cross-platform gameplay frequently considers how games can be played across multiple operating systems and hardware devices [116, 126], enabling concurrent play across Windows, macOS, iOS, Android, Xbox, and PlayStation. For example, *Minecraft* is a popular sandbox survival and building game that supports cross-platform play on PC, virtual reality (VR), and mobile devices [126]. In the context of *Genshin Impact*, an open-world action role-playing game (RPG), player perceptions of multi-platform capability suggested that cross-platform play is an affordance that enhances social experience [116]. However, communication affordances within these cross-device environments are often limited, which can negatively impact the gaming experience [119]. Specifically, PC players uniquely express a desire for in-game voice chat compared to mobile players, suggesting that PC players actively seek deeper social interaction when gaming [67, 119].

Some studies have begun to examine how multi-platform ecosystems support player interaction. Prior work has partially mapped out how Discord operates within this ecosystem. For instance, existing communities on Reddit may migrate to Discord as a supplementary communication channel [13]. Similarly, players in VR contexts used Discord to deepen relationships formed in-game and maintain community cohesion over time [2], while participation on Discord during competitive online play has been shown to support social inclusion by facilitating a collective identity [148]. However, little is known about how youth engage with multi-platform gaming ecosystems beyond individual game contexts. We examine teenagers' social interactions on Discord when gaming to understand how the broader platform ecosystem shapes their experiences.

### 3.3 Gaming Risks and Safety

Safety research defines risk as the likelihood of experiencing harm, while recognizing that contextual factors such as identity and exposure can increase both the probability and severity of harm [96, 114]. Risky interactions could cause harm through harassment, toxicity, spam, violence, sexual and financial exploitation [8, 11, 139, 150, 151]. Beyond a single platform, risks also emerge in the usage of multiple platforms. When two platforms are located in a physical space together, one platform can cause another platform to respond in a hazardous way [28].

In a multi-platform gaming ecosystem, user privacy and trust is a growing concern [122]. Using of multiple platforms can create safety risks for children, where they experience sexual violence, financial fraud and participate in illegal behaviors [45]. As platforms create an online space and facilitate social interactions [107] that can potentially be risky [76], platform governance refers to the policies and practices that manage these interactions [50, 100]. Platforms may avoid legal liability when the conditions leading to harm involve interactions across multiple platforms [31]. Challenges with policymaking and policy enforcement prevented safe gameplay, which exposed young players to harmful game designs like gore and inappropriate roleplay [150].

Users in a multi-platform ecosystem can be exploited because the movement between platforms is frequently overlooked [58]. Moreover, malicious actors in gaming spaces may leverage trust as an entry point before migrating players to external sites [36]. This occurred when scammers advertised NFT sales to lure in users on Twitter, then moved transaction discussions to Discord, where they conducted fraudulent transactions [115]. Additionally, Discord has been used to bypass governance on gaming platforms. Players could seek out harmful gaming experiences through Discord [150] and malicious actors recruited children from Discord into harmful online spaces hosted on Roblox [147].

Little is known about how youth engage in gaming-adjacent social interactions on Discord or how they are protected in these spaces. Prior work has identified toxicity as a primary driver of harm for youth [80, 96] and gamers [14, 72, 151] noting that social affordances like voice chat can expose marginalized players to gendered risks that require constant management [77, 90]. Our research advances the understanding of gaming risks and safety by investigating teenagers' experiences while gaming together on Discord.

## 4 METHODOLOGY

We conducted a qualitative study to explore how Discord impacted teenagers' gaming experiences. Approval from our institution's Institutional Review Board (IRB) was obtained before proceeding with Data Collection, which occurred from June to July 2025. We completed 16 semi-structured interviews and analyzed transcripts using Reflexive Thematic Analysis (RTA) [22].

### 4.1 Data Collection

Our data collection procedures safeguarded minors' privacy, emphasized informed consent, and reduced potential risks associated with researcher-participant interactions. Consistent with our institution's IRB guidelines, participant recruitment was limited to public-facing channels accessible to teenagers or directed toward parents or guardians who could express interest on a child's behalf. Recruitment forms were designed to capture of essential personally identifiable information only: participants' dates of birth, the legal names of both the participant and parent or guardian, and parent contact information.

We posted recruitment messages seeking teenagers (aged 13-17) who live in the United States and use Discord through multiple social media sites: Facebook, X, Instagram, LinkedIn, local public spaces, Discord servers (local, University, and teen-centric Discord servers) and Reddit (r/AskTeens and r/Teenagers). For snowball sampling, we contacted 2 parent participants via email, providing them with our recruitment materials to distribute within their social networks to parents of eligible teenagers. From online recruiting, we received a total of 24 responses, however 5 did not respond. Further, 3 teenagers were determined ineligible because they were unable to recall any gaming-related risky interactions that they experienced on Discord, even after follow-up probing. After eligibility questioning, 16 teenagers were invited to interview. The resulting participants were recruited from these platforms: snowball sampling (n=6), Viva Engage (n=5), X (n=3), Nextdoor (n=1), and locally from a flier (n=1). Participants ranged in age from 14 to 17 years, with a mean age of 15. Game titles reported by participants included: Roblox (n=8), Minecraft (n=4), Valorant (n=4), League of Legends (n=2), Clash Royale (n=1), Fortnite (n=1), Rust (n=1). Participants described playing games often on the weekends. Their gaming partners comprised of friends from school, friends who they met online but met in person, strictly online friends, and unfamiliar online teammates found from Discord. Most participants reported living in suburban areas (n=10), then urban (n=5), and least represented is rural (n=1). The demographics table is provided in Appendix A1.

The lead researcher contacted parents using the provided contact information to arrange Zoom meetings for a consent process and eligibility screening with teenagers. Parental consent was required because participants were minors. Zoom sessions were held with parents and teenagers present to explain the study and obtain verbal consent. While parental awareness may influence disclosure, the study balanced ethical obligations with the need to foster open participation by emphasizing confidentiality, conducting interviews in participant-selected private settings, and clarifying that parents would not have access to interview content. After consenting, parents were asked to leave to ensure teenagers could respond openly without parental influence. Eligibility review was conducted to understand if the teenager met two requirements: 1) a current Discord user and 2) have experienced risky social interactions through Discord while gaming.

The first criterion of being a current user was verified by requesting the teenager to share their Discord profile. Account creation dates were checked to confirm they had joined the platform before recruitment. During this process, the lead researcher seemed to build rapport with the teenager by sharing her profile and demonstrating her use of the platform.

The second criterion of having experienced risky social interactions through Discord while gaming was determined by teenagers' own perceptions of risk and harm. By allowing participants to ultimately decide which interactions were risky to them, our definition of risky social interactions was guided by their experience. We were careful to prompt the participant to share their experiences with gaming and Discord without feeling judged for their behaviors. Our semi-structured interview guide with open-ended questions is provided in Appendix A2.

Participant assent was obtained before beginning interviews, which were recorded and transcribed via Zoom. Participants were informed that their responses would remain confidential and that no identifying information would be reported. Participation was voluntary, and teens could skip questions or withdraw at any time. All interviews were conducted by the first author. She practiced reflexivity throughout the interviews by critically reflecting on her own experiences of gaming with others on Discord, consciously balancing both insider and outsider perspectives [143]. Taking notes during the interviews enabled her to identify and challenge assumptions arising from her insider position, while also prompting her to probe more deeply into participants' perceived and shared norms and experiences [54].

Throughout data collection, the first author prioritized participants' emotional well-being and observed engaged, candid participation across interviews. Interviews lasted between 22 and 48 minutes, with a mean duration of 35 minutes. A $25 Amazon gift card was sent to each participant upon completion of the interview. Audio recordings and transcripts from the interviews were securely stored on password-protected systems with access limited to the research team and were deleted after analysis. To protect participants' privacy, teenagers are referenced using identifiers (P1, P2, ...).

### 4.2 Data Analysis

We analyzed each transcript using Reflexive Thematic Analysis (RTA) to share how Discord impacted teenagers' social experiences when gaming. By following the six steps of RTA [19, 22], we used manual inductive analysis to meaningfully capture patterns in participants' shared experiences [22]. The first author independently reviewed transcripts and coded the quotes as a single coder according to the research question using semantic and latent coding [21], while being reflexive by interpreting the quotes through her own lived experiences as a Discord user and gamer. The first author also repeatedly revisited transcripts throughout the initial coding stage, which resulted in 127 unique codes. Once initial coding concluded, the research team iteratively determined coding consistency by reviewing the codes produced by the first author and discussing the meaning of original quotes to identify disagreements in multiple sessions. After finalizing the codes together, the lead author used Miro as a digital whiteboard to arrange and share initial themes with the research team.

The reviewing and developing of themes involved multiple iterations combining inductive and deductive reasoning, where the research team continuously revisited the quotes and codes to refine their interpretations. The initial themes were continuously discussed by the research team, iteratively rearranging codes when identifying patterns between codes, which is consistent with RTA [21]. Multiple discussions took place, where each researcher iteratively discussed potential subthemes and themes while providing suggestions to surface deeper, meaningful patterns [22, 104]. For instance, an initial subtheme was named "Building Community Safety Through Personal Responsibility," however, upon discussion between the research team, it was determined this title lacked clarity and did not precisely surface the meaningful interactions that participants described. Thus, the team refined the idea by proposing new theme names after revisiting the other subthemes, identifying the meaningful and resonating ideas about player resistance and moderation in gaming. Therefore, the final subtheme was called: "Resisting Player Abuse Being Undetected by Discord and Games Moderation." Each finalized subtheme and theme was validated by the research team's active engagement with their lived experiences and assumptions, aligned with RTA [21]. Afterwards, representative quotes were selected through collaborative review to determine which quotes best reflect each theme [22].

As RTA is an organic and unstructured process, we prioritized multiple rounds of interpretive discussion and iterating until final themes were generated by the research team [20]. After achieving consensus, the team arrived at a cohesive thematic structure consisting of two overarching themes: 1) Discord Cultivates Cross-Platform Gaming Collaboration and 2) Discord Exposes Teen Players to Risks While Gaming. The themes were finalized after confirming internal homogeneity and external heterogeneity [104]. After the research team reached consensus on the final themes and illustrative quotes, the first author developed a thematic map on Miro. Given the fundamentally interpretive and iterative nature of RTA, as well as the emphasis on reflexive sense-making, the calculation of interrater reliability was not considered appropriate for this analysis [97]. The following quote and analysis demonstrate the lead researcher's reflexivity and approach to RTA:

> "I made a friend in a server, in a game I was playing. The friend was in the server like someone I hadn't met before, and we started talking. But it wasn't personal information. It was just trading and stuff on the game..." P8

First, she felt connected with the participant because they described making a friend while playing a game and reflected on her own online friend-making experiences. Therefore, the initial code became: 'Felt like they made a friend when playing game together but didn't talk about personal info.' After looking through other initial codes across multiple interviews, a group of codes became especially meaningful to her because she shared a challenging social experience with finding a gaming community on Discord. This group of codes collectively described how participants felt about connecting with others through gaming. Thus, these were grouped into a higher-level code: 'Gaming connections are not necessarily intimate.' Through iterative review of many higher-level codes, she observed a meaningful pattern in participants' use of Discord to establish and sustain connections for gaming cooperation, becoming the subtheme: 'Cooperative Gameplay-Oriented Social Interaction.' The final theme was established after analyzing and interpreting multiple subthemes through continuous discussion with the research team to define a meaningful, central concept: 'Discord Cultivates Cross-Platform Gaming Collaboration.'

### 4.3 Researcher Positionality Statement

The lead researcher is a lifelong gamer and an active user of Discord. She started using Discord when she wanted to talk to friends while playing a multiplayer game that did not support voice communications. Since then, she experienced many different gaming-related interactions on Discord that informed her curiosity about the platform. Despite her outside position as a researcher, her insider experience allowed her to relate meaningfully to the participants' stories and understandings of Discord activities [12]. The second and third author have used Discord for professional communication.

Throughout analysis, we remained attentive to how participants' accounts of risk may be shaped by their backgrounds and by the normalization of such experiences within everyday gaming and trading practices. We therefore approach their narratives with caution, seeking to avoid both trivializing harm and over-interpreting participants' awareness or articulation of risk in ways that may not align with their own framings.

### 4.4 Ethical Considerations

As researchers working with teenage participants engaged in cross-platform gaming environments, we are mindful of the ethical implications of studying minors whose everyday social and gaming practices involve exposure to online risks. This includes reflecting on the asymmetry of power between researchers and participants, particularly when participants describe experiences of harm while gaming or using Discord. To help reduce this distance, the first author engaged in relevant self-disclosure by sharing her Discord presence in a manner intended to support rapport-building and participant comfort [12]. Prior to beginning each interview, she also carefully considered participants' privacy by confirming that they were in a safe and comfortable setting to discuss their gaming experiences on Discord. These practices reflect our commitment to ethically engaging with teens' gaming and Discord experiences.

## 5 FINDINGS

Our findings show that Discord plays a dual role in teen gaming experiences, simultaneously cultivating cross-platform collaboration and exposing players to gaming risks. As a persistent communication space, Discord keeps gaming engagement active beyond moments of play by supporting in-game coordination, real-time information sharing, and access to game-related resources. At the same time, the extension of gameplay into Discord's communication channels introduces risks, as player abuse on Discord bypasses in-game moderation systems. Therefore, teen gamers navigate an ongoing tension: they actively sustain social ties and collaborative play, while adapting their behaviors to manage risks that emerge across platforms. Together, these findings illustrate how cross-platform gaming participation benefits gameplay, improves socializing, and sustains risk. Our findings are summarized in Table 1.

### 5.1 Discord Cultivates Cross-Platform Gaming Collaboration

Discord in cross-platform gaming cultivates player collaboration by providing informational resources and supporting ongoing social engagement beyond gameplay. Teenagers prioritized gameplay interactions over building closeness, participating collectively in gaming activities, accessing timely notifications about in-game opportunities. Concurrently, participants also used Discord to enhance social gaming interactions, such as sustaining friendships from school and connecting with broader gaming communities. Participants were engaged across platforms, allowing gaming experiences to persist even when active play has stopped.

#### *5.1.1 Prioritizing Gameplay Over Personal Connection*

Many participants facilitating game-related communication over Discord, while keeping personal distance. This was done by intentionally keeping their Discord interactions focused on cooperative outcomes, prioritizing functional coordination over personal disclosure while gaming. P1, an avid *Valorant* player, described how she used Discord to find teammates to play games together. While this meant she was talking to strangers about pertinent in-game information, gaming with strangers was impersonal.

> "When I'm playing games [and talking to others through Discord]. It doesn't really hit me that it's a stranger sometimes, cause some people don't really strike a conversation. So, there's better comms rather than me feeling too uncomfortable... it just didn't get that deep." P1

In this context, P1's uncomfortable feeling came from becoming too personal with strangers by engaging in casual conversation outside the game. To navigate around uncomfortable feelings and a potentially risky experience, P1 used Discord strictly as a platform that allowed for clearer communication and enhanced teamwork, rather than initiating personal connection through conversation. Teamwork with strangers in-game could be facilitated through voice chat on Discord using "comms" or game-related communication, conducted for rapid exchange of critical information. For P1, using Discord voice channels only to share game knowledge and avoiding deeper conversations served a dual purpose, where she was able to benefit from enhanced gameplay while also minimizing risk exposure from strangers. Similarly, other participants' engagement with Discord was integral to multiplayer gameplay, rather than socializing.

As he explained in relation to *Valorant* and *Rust*, communication with strangers was a requirement of multiplayer games, rather than an opportunity for relationship-building.

> "I have to communicate with the strangers on my team, and it's a huge part of the game. When I started playing Rust, it was the same deal. I wasn't really talking to people on Discord and when I did, I used 'Push to Talk.' I only speak when I really have to communicate in the game. It was just communicating. Saying 'There's a person over there' or 'I died. My body is in this spot.'" P15

Push to Talk on Discord is voice setting that only transmits input audio when a designated key is pressed [154]. This setting enabled P15's highly selective participation, limiting his speech to moments where game-relevant information needed to be conveyed. This selective and task-oriented use of communication highlights how platform features can be mobilized to maintain gameplay-oriented interactions. By keeping communication narrowly focused on competitive coordination, P15 minimized opportunities for personal or social exchange, effectively using technological affordances to manage both performance and exposure when interacting with unfamiliar players. To improve competitive aspects of gameplay, there is also potential to form connections through learning and teaching game concepts. P9 maintained personal distance with an online stranger by grounding their interactions in a teaching dynamic, where they learned about each other's gaming approaches and preferences.

> "It's teaching how to play games as like a tutor for characters I play. And I just teach normally. So, it's never talking about your own life. It's more talking about the game." P9

For competitive gamers, like P9, they use Discord servers to share their screens that display their gameplay and review strategies. P9 maintained professional and informative boundaries, not extending discussions into personal life. He used this social boundary within teaching gameplay to other players to cultivate cooperative relationships with unfamiliar players. Another participant emphasized that enjoying gameplay was more important than developing personal relationships when interacting with online strangers. To focus on having fun while gaming on Discord, P16, a 14-year-old, explained that she "usually deters people" by disclosing her age early in conversations, as this helped her avoid "weird and romantic" interactions. In her experience, when others knew her age, it often discouraged others from pursuing those types of conversations. Instead, she preferred relationships centered around shared gaming activities, describing one friendship positively because they primarily played games together on Discord: "I really like playing games with them as long as they're not being weird to me in any way." Despite a 9-year age gap between P16 and her online friend, "sexual or romantic" conversations were considered unacceptable and focused on gameplay rather than personal intimacy. By centering communication with others on Discord towards gameplay and skill development, players maintained clear relational boundaries when fostering collaboration. Overall, participants leveraged Discord for cross-platform cooperative gameplay while maintaining social boundaries that minimize personal disclosure and risky interactions.

#### *5.1.2 Staying Informed about Games through Discord Servers*

Participants described using Discord as a central hub for information, where they joined servers to review game updates, learn about competitive communities, and acquire in-game resources. This supported collaboration because timely information about the game encouraged them to stay engaged with their gaming communities. For teenagers like P6, game-focused servers operated as informational hubs that supported ongoing awareness of game updates.

> "Right now, I'm in a server and they would have announcements and updates, and everything in the game that's new, and they would tell me about it. And then that would lead me to using it more." P6

As P6's account illustrates, announcements and update channels structured how new content was encountered on Discord. Moreover, ongoing server membership sustained awareness of gaming updates, which appeared to encourage P6 to engage more frequently with both the platform and gameplay. This suggests that Discord servers are leveraged to sustain player engagement with both the community platform and the game. Moreover, some participants described being members of multiple Discord servers. Their membership in multiple servers allowed them to benefit from specific information that was offered by a particular server. P4 described the purpose of some large servers were promote other servers, encouraging membership in multiple communities.

> "There are these types of servers where they would have links to other servers. People would promote their servers and it would just be like there'd be descriptions of what the servers were about, and so I would join the ones that interested me… There was an *Animal Crossing* server. I was a huge *Animal Crossing* player." P4

This kind of server broadened awareness by acting as an information hub for other servers, introducing and connecting its members to similar gaming communities. In P4's case, the specific description of a server let her become aware that there was an *Animal Crossing* server that she could join and gain access to their game community resources. P11 found that it was useful to be a member of multiple servers because they provided her information through real-time notifications outside of the game. She had been playing *Grow a Garden*, a popular game on *Roblox* [51], where players manage a virtual plot of land and buy plant seeds from a shop that refreshes with a random selection of seeds every five minutes.

> "Discord servers tell you what's in stock, since the stock changes and everyone's waiting for good items. I keep track using my phone because you can choose to get notifications based off what's in stock… It's also helpful because then I don't have to stay in the game for a bunch of hours, just waiting for something to be in stock." P11

For players such as P11, Discord notifications supported her attention to timely updates in-game, alerting her about opportunities to acquire specific in-game items (such as seeds for *Grow a Garden*) when they became available for a limited time. This allowed her to remain engaged with the game without needing to maintain continuous in-game presence. Additionally, P11 believed that manual checking of the shop for rare items was an inefficient use of time. By adopting Discord notification features, players like P11 were able to manage their attention and time more strategically, positioning the platform as a tool for optimizing gaming engagement rather than sustaining constant gameplay. In contrast, some participants described more sustained forms of server engagement.

> "I'm probably checking my servers almost every day just to see if there's any updates or game-related codes and things like that. There's usually a lot of people... maybe a couple 1,000 and the servers are pretty organized. Really straight to the point about codes and updates." P8

As P8 explained, he monitored Discord servers almost daily to remain informed about game announcements, updates, and in-game codes. This reflects how players may rely on Discord to access timely information and maximize gaming participation opportunities. The scale and organization of these large servers were salient to him, with their structured channels and concise information enabling efficient monitoring. P8 also noted that other members engaged in similar practices, highlighting how routine server checking became a normalized way of staying connected to ongoing game activity. Across participants' experiences, they valued Discord servers for providing timely information beyond the game.

#### *5.1.3 Sustaining Offline Social Connections through Gaming Together on Discord*

Players used the affordances of Discord to extend and sustain offline relationships through coordinated gaming activities across platforms. Through chatting, coordinating gameplay, spectating matches, and exploring shared game interests, participants reinforced existing social bonds with friends while maintaining ongoing participation in shared gaming experiences. P15 explained that when playing Minecraft with friends from school, voice communication was an expected and integral part of their shared gaming experience.

> "I turned 13 and I had recently switched to PC… I needed a new means of communication between my friends, because we used to play on Xbox with the game audio on Speaker and a Zoom call. But I couldn't use Zoom on my laptop while playing Minecraft." P15

This illustrates how changes in device use can necessitate new communication platforms. As P15 transitioned to PC gaming, previously used communication methods no longer fit their gaming setup, prompting the adoption of Discord to maintain voice communication with friends. In this way, Discord filled a gap created by device limitations, enabling coordinated communication without disrupting gameplay. Further, many participants spend a lot of time in voice channels on Discord while gaming as a means of socialization with offline peers. Several participants explained that voice channels functioned as open, persistent spaces where members could enter or leave at will. P2 described interacting with his friends by talking to them in voice channels (VC) while they were gaming to build familiarity.

> "I usually talk to them every day, because they're online or playing their game. I just come and watch their stream… They're usually in VC so I just join. And I'm like 'Hi!' and then we talk for a few hours… I became really good friends with these people because I interacted with them so much on Discord and we just talk in VC." P2

For P2, this created a low-pressure environment for interaction. Discord's server design provides visibility into who is present in a voice channel and what is happening, allowing members to join ongoing conversations or activities as they choose. In P2's case, he would see when other members were streaming gameplay and could join their VC to socialize. Participants like P2 were able to gradually develop friendships by becoming familiar with others through VC. Beyond sustaining offline friendships, P1's account suggests that Discord itself operates as a marker of gaming identity. In the process of making friends, teenagers like P1 would exchange social media connections.

> "Most of the friends I made at school also use Discord [and they] are also gamers. [If] I know they play games, they probably have a Discord. But if they don't play games. I'll ask for their Instagram… I would say Discord is my main form of communication, because most of my friends use it as their main platform too." P1

P1 recognized that Discord functions as the primary social platform for gamers in her community, allowing her to distinguish peers who share gaming interests from those who do not. In this way, the presence of Discord itself signals membership in a gaming community, helping participants like P1 identify and connect with potential friends. Across participants' experiences, collaborative gaming on Discord supported both continued social connection and coordinated interaction beyond face-to-face settings.

#### *5.1.4 Building Discord Gaming Community Connections Online*

Discord provides an online space for players to build community and connection while playing games. Participants described using Discord to locate gaming communities centered on shared interests, connecting with others who played the same games. Even with a wide spectrum of interests, they were able to find communities that aligned with their preferences, regardless of whether the games were popular or niche. These spaces fostered a sense of connection and belonging by bringing together players around shared gameplay goals, including competitive play. For example, P15 "was a top leaderboard player for a while" and desired "to get into competitive lobbies or talk to other top players and get into parties with them." Therefore, he used Discord to join a variety of competitive servers and meet other passionate players.

> "I'd add people and join servers… competitive communities. I met a couple people that were normal and agreed with the things I agree with… I enjoyed playing with them more than other random people. I got bored of playing in random teams…And then it's I kept kind of meeting more people… they're actually nice people that I'll continue to play with." P15

P15's account illustrates a preference for stable, predictable team interactions over randomly assigned teammates, which he found less engaging. Discord facilitates this by connecting teenagers like P15 to competitive gaming communities, transforming gameplay into a socially meaningful experience. Once integrated into these communities, he actively sought out more personal connections with teammates, extending collaboration and sustaining shared gameplay over time through more permanent teams. Although many participants often used Discord for social gaming, P2 was focused on expanding his competitive community connections for networking purposes.

> "I'm a very high rank. I want to potentially broaden my [professional] career in the game. I was like, 'Where could I meet more people or the communities of what I do' and that's how I found Discord... It's been more helpful than not helpful." P2

For P2, using Discord to connect with other competitive players not only provided social engagement but also potential career opportunities in gaming. This shows how teenagers like P2 could develop professional gaming connections with competitive players, extending gaming connections beyond play. Moreover, P8's experience with building gaming connections shows how the initiation of gaming connections is not confined to a single platform but instead emerges fluidly across game and non-game spaces.

> "Depending on what server I found them from, I send them updates or things I found on the game that could be fun or something new. And if [I met them] from a Roblox game, I might send game suggestions of other games." P8

Initial encounters may occur within games or on Discord servers, but efforts to build and sustain gaming relationships often extend beyond those original contexts. P8 recognized where connections were first established yet deliberately maintained them on Discord, illustrating how the platform functions as a persistent social infrastructure independent of active gameplay. Its chat channels enable coordination, socialization, and collaboration even when players are offline or playing different titles. However, some participants preferred to maintain gaming connections without continuous interaction outside gameplay. For them, relationships developed through shared community participation, observing others, and communicating only when comfortable. P7 described a cautious approach, gradually developing familiarity by listening to group conversations on Discord while playing *Minecraft*.

> "I didn't use voice chat for the first 2 months... I used Minecraft books in-game and I'd throw them at people whenever I wanted to say something. It was a running joke that I'd do this. I just listened to them talk about their lives until it felt like I actually knew them [then I started talking]." P7

P7 adopted an in-game feature in *Minecraft* that allows players to write messages in books and share them with others [47], using this mechanic to routinely and indirectly communicate with other players without relying on Discord's text or voice channels. This shows how Discord could facilitate parallel participation, allowing players to communicate while playing together without requiring immediate or reciprocal self-disclosure—particularly for games without built-in voice chat. Across participants' experiences, they intentionally sought out relatable gameplay contexts to build a gaming community, highlighting how Discord servers support social engagement around gaming

## 5.2 Discord Exposes Teen Players to Risks While Gaming

Discord simultaneously supports gaming participation and introduces heightened risks that require ongoing negotiation. Discord communication channels allow harmful activity to bypass in-game safeguards and is used to facilitate the spread of game-related scams. When seeking social experiences related to gaming, teens were exposed to various risky interactions with their gaming partners. Moreover, participants needed to overcome limitations in platform risk moderation because player interactions took place across games and Discord. Rather than disengaging, participants remained active on Discord by adapting their behaviors to reduce harm. Together, these dynamics illustrate how teens balance the benefits of staying connected to gaming communities with the risks that emerge as gameplay extends into persistent social spaces.

### *5.2.1 Subverting Game Limitations through Discord Creates Risky Situations*

Game limitations manifest as restrictions on in-game voice communication and as the absence of certain gameplay mechanics. Consequently, many participants reported turning to Discord as an alternative platform that better accommodated their preferred modes of social interaction. For P16, using Discord offered her the freedom of expression without being limited by profanity filters.

> "In Roblox voice chat, you're limited with what you can say, because obviously it's a kid's game. If you say a bad word, you get VC banned and whatnot. With Discord, you can communicate with more people and you can call freely. So it's just funner and easier to communicate." P16

Swearing in voice chat conflicts with the norms of kid-friendly games. Thus, P16 understood that getting auto moderated for using profanity in Roblox was a common occurrence and adopted Discord to avoid being punished. However, this becomes risky as she deliberately engages in unrestricted conversations on Discord. Some participants also identified Minecraft as a game with limited communication affordances, noting that the absence of native voice chat prompted them to rely on Discord to facilitate real-time interaction. Therefore, players frequently invite others to join Discord servers for voice communication. P12 was looking to join a Discord server to voice chat with other players when he came across something that appeared to be a valid Discord server.

> "You click the link and it looks like a regular Discord server… but once you're in the server, they ask you to verify a code sent to your Windows account, saying it's from an official bot. But it's a different bot that sends your information to the hackers who own the server. They end up taking your Minecraft and Discord accounts and use those hacked accounts to keep spreading the scam." P12

When attempting to enhance gameplay communication on *Minecraft*, P12 encountered a risky, fraudulent design that abuses players through account takeover. These compromised accounts are subsequently leveraged to propagate the same malicious invites, indicating a self-perpetuating account takeover campaign. P12's experience with this risky activity illustrates how subverting in-game communication limitations by shifting interactions to Discord channels can introduce significant security risks. Furthermore, user trust in Discord's invitation system can be exploited, demonstrating how external communication platforms can amplify risk when integrated into gaming. Other limitations reflect gameplay features that are simply not supported within the game environment. Multiple participants described leveraging Discord to move communication off platform to participate in a player-created trading practice that existed outside the game's official design. P11 shared how this trading practice led to her being scammed.

> "For games where you can't really trade, you have to gift something and then they'll give you something back. Roblox has a rule against that, but a bunch of people do it [as] Discord conversations. They might say, 'You can trust me.' Then they'll give you their Discord… [Having] their Discord, I have like another connection to them, so it's more trustworthy." P11

Roblox had a rule against "trust trading," where players use a game's gifting function to exchange items. This method of trading introduces risks for players because it relies on the other, typically unfamiliar, player to reciprocate. Additionally, the coordination of trust trades through Discord also heightened P11's sense of trust in others because sharing profiles increased their sense of familiarity. Many participants described how they encountered risky situations on Discord when gameplay interactions were moved outside the game's moderated environment. Participants explained that game trades often resulted in scams, where the trusted player does not follow through with their end of the trade. P6 had been scammed before when she was looking for trades on a large Discord server for a Roblox game, explaining that suspicious players put up trade requests.

> "[Scammers] would put a really good trade like no one would ever do this trade. And then people would fall for it… they would be like, 'Oh, my God! That's such a good trade. I have that item.' And then they would 'Log into this website and do the trade. I'm online right now.' And then they scam you. I didn't know it was or could be a scam, because it literally looked exactly like the app." P6

Scammers leveraged "trust trading," which was made possible by Discord's features and perceived trustworthiness, to establish credibility that increased the likelihood of deception. P6 encountered a risky trade request on Discord in which a scammer altered standard trading procedures by redirecting her to a phishing website disguised as the legitimate platform. This experience illustrates how deceptive designs can prevent players from recognizing fraudulent game sites, leading them to unknowingly disclose account credentials. Although Discord enhances coordination and social engagement, it increases teenagers' exposure to social and gaming risks.

#### *5.2.2 Encountering Risky Experiences During Gameplay Chats*

Although many players have positive experiences with gaming together on Discord, several participants encountered risky situations when chatting. Multiple participants described how relationships formed through gameplay could become risky as gaming and socializing blended. Although P1 enjoyed playing Valorant with friends, she recalled how it felt risky to play games with high competitive stakes because it introduced toxicity into her group.

> "[We were] playing a game and our whole team lost 13 to 0. We lose a billion RR. They get pretty sad or mad… At a certain point we had a falling out because people started bringing up personal matters when they got mad… during that Valorant game they got upset and used personal insults… too heated in game and then they yell at each other." P1

When playing competitive games, conflicts between friendly players can escalate and result in toxic behavior. P1's experience with hostile exchanges among friends that ultimately resulted in the dissolution of a friendship demonstrates how competitive pressure can undermine social dynamics within gaming environments. This illustrates how players are required to simultaneously manage gameplay performance, emotional regulation, and peer conflict during Discord calls, with failures in this balance leading to interpersonal harm. Moreover, a few participants described how friendships centered around play could unexpectedly lead to inappropriate social situations, such as exposure to harmful content or emotionally distressing disclosures. P7 regularly interacted with a girl on voice chat in Discord while playing *Minecraft* together and built a close friendship by talking about their personal lives.

> "She talked about her in real-life problems… She had a boyfriend that sexually assaulted her. She got pretty graphic… describing how she was trying to scrub away at her skin to make her feel better because she felt like he was still touching her. I was 12. It was jarring because we just played Minecraft together… I had never dealt with anything like that before I was on Discord." P7

P7's friend confided in her about her traumatic sexual experiences, which made her feel extremely shocked and unprepared. As a younger player, P7 did not know how to appropriately respond or support her friend, but she was empathetic to her friend and tried her best to support her. This shows how online friendships can blur personal boundaries and expose teens to distressing content beyond their developmental capacity. Furthermore, participants reported that social connections from gaming often expanded through friendly introductions to unfamiliar individuals. P16 described becoming acquainted with some unfamiliar individuals after being invited to voice chat on Discord while she was gaming with friends.

> "People are 'freaky' in a joking way and I sometimes do it back. But [not] romantically… I can tell when they're joking and not joking… I have some friends who people have sent genital pictures to them and they send some stuff back. It's very lewd. As long as my friends are nice to me, I couldn't care less what they do. It's their life, their decisions." P16

The gaming community P16 was part of had a norm of unsafe or inappropriate behaviors. While she did not participate in the risky behaviors herself, the exposure occurred because risky themes were part of that server's entertaining conversations. When referring to "freaky," P16 is describing openly engaging with risky themes, such as sexual innuendos or sexualized flirting. Her experience indicates that while gameplay remains central, socializing while gaming may expose teens to risky themes because there are no limitations on discussed topics. Overall, participants' experiences underscore how conversations on Discord while gaming can rapidly shift from gameplay to inappropriate issues.

#### *5.2.3 Resisting Undetected Abuse Across Discord and Games*

Player abuse resistance occurred because neither Discord nor game platforms effectively moderated abusive behavior or enforced accountability. Multiple participants recognized that when risks span private messages and multiple gaming platforms, moderation mechanisms from both game environments and Discord servers are limited, often allowing abusive or fraudulent behavior to go unnoticed and remain unresolved. While some moderators issued warnings or removed offending users, P15 described inconsistent feedback across platforms, noting that "in some games, if you report someone you get a notification later, but on Discord, you never know if they get banned," and expressed disappointment that "you don't get any follow up ever." As a result, participants like P12 assumed responsibility for sharing information about suspicious activity and warning newcomers.

> "It's hard to report people for spreading [scams] because they don't always post it in chat. There isn't a specific report option… they put the link on signs, so [moderators] can't see it. There's usually a message like 'permanent party for this game,' and when people join, it warps them to the person's island. They don't say anything, so it doesn't look suspicious, but there's a sign on the wall that says something like 'join the Discord server for VC.' Sometimes I warn people that it's a scam and then the scammers disband the party right after." P12

In P12's experience, abusive interactions often begin within the game and then shift users to external Discord links, illustrating how the multi-platform nature of these interactions creates oversight gaps. Game-related activities, such as trade negotiations, are frequently conducted within private Discord spaces where moderation visibility is limited. P10's experience shows how this shift in interaction context demonstrates resulted in only partial protection from Discord moderation.

> "There's a system where people will search up usernames in the scam channel to see if they've done it before. Eventually they'll probably get banned from the server if there's too many reports. Most of the time, people scam privately... Schools don't really teach you that. [But] there's a bunch of YouTube videos online about people trying to catch scammers." P10

While scammers posting suspicious links in public channels are more likely to be identified by moderators or the community, this visibility incentivizes malicious actors to shift interactions into private communications. P10 and other participants shared how they were required to independently verify the legitimacy of potential trading partners and manually report suspected scammers within dedicated community channels. Additionally, P10 further noted that strategies for navigating player abuse and scam risks were often learned informally through external online resources rather than through in-platform safeguards.

Notably, a few participants described actively adapting their behavior to resist in-game harm. Some participants further opposed scammers by refusing to disengage, which resulted in controlling the risky situation while continuing to participate in gameplay. P11 encountered a scammer who proposed a deceptive trade deal when she was on a Discord server dedicated to *Grow a Garden* game discussion. She was talking to another player within this server to trade pets when she recognized that they were attempting to scam her.

> "I know that they're a scammer so I try to waste their time like, 'Oh, yeah, I'll do the trade.' And then make the conversation go for a long time. In games like 'Grow a Garden,' I'd be like, 'Hey, could you give me 5 minutes? I need to ask someone about this trade.' But I wouldn't actually ask anyone. And then I'd come back later and I'd say, 'Oh, they want me to change this to something else.' and then I would make them join the game... It would be really funny." P11

*Grow a Garden* is a Roblox game where players can use "trust trading" to exchange items, such as plants and pets, with each other. Rather than disengaging from the scammer's trust trade, P11 considered the situation as an opportunity for entertainment by deliberately leading the scammer on and wasting their time to observe their reactions. By deliberately prolonging trade interactions and feigning compliance, she exercised control during a risky interaction without any moderator intervention. This shows how teens could assert agency and maintain safety, turning a potentially harmful exchange into a managed experience that balances risk with playfulness. Altogether, participants attempted to resist player abuse but often struggled because moderation is fragmented and no single system provided complete oversight.

Table 1: Theme Summary

| Theme Name (Participant count) | Definition | Representative Quote |
|---|---|---|
| Theme 1: Discord Cultivates Cross-Platform Gaming Collaboration (14) | | |
| Prioritizing Gameplay Over Personal Connection (8) | Players intentionally maintained social boundaries by limiting personal disclosure, emotional intimacy, and relationship development beyond gaming-related activities. | "When I'm playing games... It doesn't really hit me that it's a stranger sometimes... So, there's better comms [on Discord] rather than me feeling too uncomfortable...." P1 |
| Staying Informed About Games Through Discord Servers (7) | Players gained access to timely game-related information, coordinated gameplay participation, and remained engaged with cooperative and competitive gaming communities. | "[A server] would have announcements and updates, and everything in the game that's new." P6 |
| Sustaining Offline Social Connections Through Gaming Together On Discord (7) | Players sustained friendships through coordinated gaming activities, shared gameplay experiences, and ongoing interaction across online and offline contexts. | "I needed a new means of communication between my friends... I couldn't use Zoom on my laptop while playing Minecraft." P15 |
| Building Discord Gaming Community Connections Online (7) | Players connected with gaming communities, socialized with online strangers, and engaged in both cooperative and competitive play. | "I want to potentially broaden my career in the game... And that's how I found Discord." P2 |
| **Theme 2: Discord Exposes Teen Players to Risks While Gaming (11)** | | |
| Subverting Game Limitations through Discord Creates Risky Situations (8) | Players experienced risky social interactions when moving between moderated games and Discord. | "In Roblox voice chat... If you say a bad word, you get VC banned. With Discord, you can communicate with more people and you can call freely..." P16 |
| Encountering Risky Experiences During Gameplay Chats (8) | Players were exposed to inappropriate content and blurred boundaries between gameplay and personal communication, creating potential social and emotional risks. | "She started talking about her in real life problems... It was jarring because we just played Minecraft together." P7 |
| Resisting Undetected Abuse Across Discord and Games (8) | Players actively resist abusive behavior that goes undetected across Discord and game platforms. | "It's hard to report people because they don't always post it in chat. There isn't a specific report option... I warn people that it's a scam." P12 |

## 6 DISCUSSION

We discuss gaming together on Discord as a cross-platform practice that supports players' social needs and extended gameplay. When teenagers use Discord alongside gaming platforms, they experience support for gameplay coordination and sustained social interaction beyond the game environment. We also introduce the "platform gap" to describe governance challenges faced by players in a cross-platform ecosystem. The platform gap fragments moderation, obscures visibility of risks, and redistributes safety responsibility onto players. We argue that cross-platform gaming is not simply a gameplay enhancement, but a player experience of social connection and risk. We call for design interventions that bridge governance, improve transparency, and center player safety across platforms.

### 6.1 Gaming Together on Discord as a Cross-Platform Practice

Our research expands on cross-platform practice beyond simultaneous use of games, showing how gameplay and social experiences emerge through interconnected platforms where actions on one shape outcomes on another. Teenagers use a platform like Discord to extend specific gaming activities beyond the game and for their social needs. Rather than merely supplementing gameplay, Discord reshapes how teens coordinate play, form relationships, and navigate platform limitations.

Teens turn to Discord when gaming platforms insufficiently support their need for social connection. While prior work suggests that games can satisfy social needs through in-game communication affordances that support coordination and relationship building [35, 88, 108, 111, 129, 130], some games played by our participants offered limited communication systems. These constraints failed to meet teens' needs for relatedness. For example, P16 found Roblox's chat system too restrictive for self-expression, so she started using Discord. Although many gaming platforms provide affordances for communication (e.g., text chat, voice chat, or friend systems) [42, 42, 43, 140], teens in our study found these affordances limiting because interaction remained tied to a specific game platform and active gameplay. Overall, teenagers adopted cross-platform practices when games insufficiently supported socialization.

Cross-platform practices allowed teens to navigate extended gameplay by redistributing interaction across platform boundaries. Prior work on tabletop role-playing games describes Discord as enabling parallel participation [138]. Our findings further demonstrate that players actively manage movement between platforms, using different spaces strategically to support emerging social ties and gameplay goals. Participants selectively transitioned interactions across contexts, for example, P7 moving from Discord to in-game exchanges to accomplish collaborative goals or escalating from text chat to voice communication only after trust had developed. Communication often required participants to be logged into the same game, constraining opportunities for ongoing social connection outside play. For example, P8 used Discord to deepen friendships through game sharing and ongoing conversation, supporting social connection even when they were not actively gaming together. While research on boundary management across platforms focused on privacy and data sharing practices [141], cross-platform practices reveal how players strategically navigate platform affordances to compensate for the relational and communicative limitations of any single gaming environment.

#### *6.1.1 Discord Extends Gameplay*

Discord in cross-platform practice is shown to extend gameplay beyond a single game. Our study highlights teenagers' cross-platform gaming experiences, where they interact with others in games and on Discord at the same time, relying on textual and verbal cues to guide play. Discord is embedded in everyday gaming routines by connecting players together outside of the game environment, allowing them to participate in trading mechanisms and receive gaming information.

Gaming connections were made concurrently in-game and on Discord for exchanging in-game resources. Prior work highlighted the importance of physical presence for trading items because players in physical settings together could use bodily signals as information and inspect each other's items by viewing one another's screens [69]. In contrast, our study shows that text channels in Discord servers, not physical cues, enabled players like P6 to coordinate trades with strangers outside their immediate social circles who possess desired items. Thus, Discord facilitates opportunistic gaming connections by reducing reliance on physical presence for resource exchange.

Furthermore, Discord servers extended gameplay by becoming sources of important gaming information, offering convenient access to game announcements with minimal social interaction. While prior research has characterized Discord as a platform where community gameplay and social interaction are engaged with separately [128], our findings indicate that players integrate Discord into their gameplay practices. Participants used Discord to coordinate team strategies, track real-time updates, and sustain interaction beyond the game interface. For example, P1 relied on Discord for temporary coordination during competitive play. Rather than serving only as a supplementary communication platform, Discord became embedded in teenagers' everyday gaming routines. Their cross-platform practices supported gameplay coordination and information access that was not fully afforded by games alone.

#### *6.1.2 Discord Enhances Social Gaming*

Discord distributed communication across platform boundaries, providing a unique social gaming experience. In our study, cross-platform practices involving Discord enabled coordination, sustained interaction, and social presence. Prior work has examined social presence in VR gaming [46] and physically co-located gaming experiences [68], showing how shared environments foster connection among players. Instead, Discord provides text and voice communication channels rather than shared immersive spaces, suggesting that synchronous communication can similarly foster a strong sense of connection, even with limited personal disclosure.

Gaming relationships could emerge through sustained social interaction rather than interpersonal disclosure alone. While prior work has examined games as deliberate interventions for social skill development [57, 136, 153], our findings suggest that cross-platform gaming practices scaffold relational competencies, including coordination, mentoring, and emotional support. Participants like P7 developed trust with her gaming community through repeated gaming sessions in shared Discord spaces. Furthermore, teens became closer with strangers and friends by watching gameplay, joining calls, coordinating activities, or passively participating as spectators. Discord's persistent channels and visible activity lowered barriers to interaction, helping teens initiate conversations and gradually deepen communication around shared interests.

However, using Discord while gaming did not necessarily translate into deeper relationships. This finding complicates assumptions derived from Media Richness Theory that people use richer communication channels to promote trust, self-disclosure, and interpersonal intimacy [88]. Instead, participants facilitated gameplay coordination and maintained social presence using Discord's richer affordances (eg., voice chat and screen sharing). For example, P15 used Discord's voice communication features to coordinate gameplay in games that lacked built-in voice chat, highlighting how richer media were often employed to meet functional and collaborative needs rather than to foster greater relational closeness. Moreover, P7's experience suggests that trust emerged not from media richness of either platform, but from the interplay between gameplay and Discord's communication affordances, which gradually fostered comfort and familiarity.

Our findings suggest that cross-platform gaming practices allow teen gamers to routinely engage in practices that span games and external platforms. This furthers prior work that has primarily examined player behavior and game design within individual platforms and game contexts [1, 23, 86]. Platform studies discursively frame users as empowered public broadcasters to promote a populist ideal [48]. Instead of promoting their personal ideas, teenage gamers utilize Discord as a platform to strategically manage their social and gameplay needs, providing functional coordination with other players. While cross-platform practices may fulfill social needs and extend gameplay, these practices may also expose players to risks that are not fully governed by any single platform.

### 6.2 Teen Online Safety in the Gaming Platform Gap

The "platform gap" represents a fundamental breakdown in platform governance, where an activity is split across distinct platforms, creating a blind spot in which moderation policies do not overlap. Platform governance is designed around the assumption that platforms can observe, regulate, and enforce behavior within their own technical boundaries [50]. However, when transactional practices span multiple platforms, as in Discord-mediated Roblox trading, responsibility and visibility become fragmented across platforms. These conditions create the platform gap, where harmful interactions can be initiated in one environment and produce consequences in another. While prior literature has examined structural friction in cross-platform design [25, 106], the challenges of multi-device ecosystems [6, 15, 93], and how users navigate polymedia to manage relationships [91], the platform gap arises from fragmented governance and the safety tensions created when activities span multiple platforms. This differs from networked governance, which focuses on coordinating moderation across separate communities within a single platform [115]. Instead, the gaming platform gap highlights how using multiple platforms together can introduce unique risks to teen online safety. Specifically, our findings reveal gaming and security risks that teens are responsible for managing on their own.

In the gaming platform gap, the responsibility of safety is placed on players. Our study shows teens taking responsibility for their safety while experiencing fraudulent trading activities. For instance, P10 took responsibility for identifying scams by consulting public resources on Discord. Building on prior work showing that players may encounter risks through Discord and other gaming-related spaces [150], our findings highlight how teens also develop self-protective strategies when platform governance is fragmented. While standard in-game systems mitigate harms like toxicity through automated blocking and reporting [73, 77, 152], cross-platform gaming shifts greater responsibility onto teens because interactions move across spaces with limited coordination between platforms. Prior work has shown that gaming toxicity can follow players across platforms [77, 95]. Our findings broaden this understanding by showing that cross-platform harms also include security-related risks, where teens must independently assess trustworthiness, identify potential threats, and protect themselves without consistent oversight or intervention.

Fraudulent activity, including trade scams and account takeovers, unfolds dynamically through cross-platform interactions. While players in dedicated trade groups are known to encounter scams [78], our findings show how these risks materialize across a gaming platform gap as reciprocity and trust-building become mechanisms for producing security risks rather than solely supporting social exchange. Although prior research shows that players engage in reciprocal activities, like trading scarce resources or exchanging virtual gifts, to express appreciation [16, 27], scammers actively appropriate these same social mechanisms. Consequently, security risks emerge through the direct interaction between Discord-based transactional communication and in-game resource exchange. For example, P11 used Discord to establish trust with potential trading partners before completing in-game exchanges, illustrating how cross-platform coordination becomes integral to transactional play and structurally produces risk.

While platform governance intervenes through platform-specific rules, technical affordances, and moderation systems [49, 50], these mechanisms have limited capacity to address harms distributed across platforms. Malicious actors exploit the platform gap by using both gaming and communication platforms to conceal and carry out scams. For instance, scammers may advertise highly favorable trading conditions on Discord to persuade participants, such as P6, to engage in fraudulent trades. Unlike scalable attacks relying on templated deception like mimicking legitimate website designs [9], attackers adapt their tactics directly to the game context. In *Minecraft*, as seen with P12, attackers embedded disguised links within in-game objects to prompt player interaction, bypass moderation systems, and redirect players to suspicious Discord servers. Our study builds upon prior work showing that individual personality traits, such as openness, can increase scam vulnerability by encouraging disclosure [53], by demonstrating how scam exposure is also shaped by cross-platform transactional events. Scams emerge not solely from individual susceptibility but through gaming practices that span platforms, where players participate in coordinated trades or respond to invitations embedded within cross-platform activities.

Harms in cross-platform gaming are socially produced within an interconnected ecosystem rather than emerging solely from individual platforms or encounters with strangers. Prior work has emphasized how young people use multiple platforms to maintain relationships and strengthen existing social ties [83, 85]. Our findings reveal a more complex dynamic: the same cross-platform practices that support social connection can also create conditions for harm within friend groups, where competition, collaboration, and relationships become intertwined. Participants like P1 used Discord to coordinate gameplay, such as discussing strategies in *Valorant*, yet these collaborative interactions could also shift into conflict. These findings illustrate how gaming tensions can spill across platforms and become embedded within broader social relationships.

Discord's affordance of persistent, out-of-game audio channels that operate entirely outside of active gameplay oversight introduces new safety challenges for teenage players. Unlike text-based communication, voice interactions are ephemeral and difficult to retrospectively audit, creating governance challenges because harmful interactions may be difficult to verify, document, or address after they occur [64]. These affordance properties create distinct governance challenges by reducing the visibility of harmful interactions. These challenges are further amplified in cross-platform practice, where voice communication occurs on Discord while gameplay takes place on *Roblox*, fragmenting oversight across separate platforms. As a result, teens cannot rely solely on platform moderation and instead assume greater responsibility for managing their own safety by exercising judgment about when and with whom to communicate. Prior work highlights communication features in games as supporting positive social interaction [135, 140] and tactical coordination among competitive adult players [125]. Adding nuance to this perspective, our findings show that teenage players adapted their communication practices to account for safety challenges. Teenage gamers in our study responded by limiting voice chat to brief, task-oriented callouts, using brevity as a self-protective strategy to reduce the risks of interacting with online strangers.

Online safety for teen gamers is shaped by a platform gap in which Discord and games operate as an interconnected ecosystem yet independently governed spaces. We introduce this concept to explain how cross-platform ecosystems structurally produce risks and shift the responsibility for safety onto young players. Despite these risks, teens continuously utilize Discord for coordination and meaningful connection, even when in-game systems restrict interaction for safety. Ultimately, our findings highlight the need for cross-platform governance that supports players' gaming experiences and safety across platform boundaries.

### 6.3 Design Implications

As players extend their gaming activities beyond the game and into social spaces on Discord, it becomes critical to ensure that cross-platform practices do not undermine their safety. Our recommendations highlight how game developers, platforms such as Discord, policymakers, families, and teenage players can more intentionally support cross-platform practices.

To improve cross-platform moderation, designs could account for layered mechanisms that enable player abuse. As illustrated by P12, scammers entice players to leave the moderated game environment by directing them to Discord, where conversations can then be shifted from public servers into private direct messages, reducing visibility and accountability. Moreover, effective moderation requires transparency to help players understand how and why decisions are made, fostering gaming justice [89]. This could mitigate negative cross-platform experiences on Discord, as P15 felt frustrated after receiving no feedback on his reports. Designs that provide personalized feedback explaining who was sanctioned and why can better inform players and strengthen their sense of safety during in-game activities. In turn, safer trading practices can enhance developers' reputations and reinforce trust in platforms such as Discord.

Cross-platform gaming safety designs should center the player experience. Players must understand the risks they encounter across platforms and feel prepared to respond to evolving threats. In some cases, this even includes strategically engaging with scammers, such as P11 leading them on for entertainment, highlighting the ways players navigate risk in cross-platform spaces. As visibility of risks becomes diminished in the platform gap, education about risks and proactive strategies become even more relevant for cross-platform protection. Recent Discord safety features face concerns with privacy at the expense of teenager safety [39].

Designs should preserve privacy of teenagers while providing them with enough access to fulfill the game activities that they want, such as socializing with friends or growing a gaming community. To safeguard socialization, studies from cyber ethics shows that social media etiquette could help to mitigate miscommunication [3, 102, 121]. Educating players about social etiquette in gaming could help them navigate difficult conversations with others, such as toxicity and conflicts with friends. While schools often struggle to keep pace with rapidly evolving online risks, families can take a more proactive approach. As P10 explained, players and parents can turn to platforms like YouTube, where content creators frequently document and expose emerging scams on Discord and within popular teenage games. Seeking out these independently developed resources could inform families about platform-specific risks to engage in more timely, ongoing conversations about online safety. As prior work has shown that public blacklists have limited effectiveness against account takeovers and scams [9], this strategy moves beyond reactive reputation systems toward a more proactive learning approach.

Designing clearer signals of players' gaming intentions could foster trust and transparency between players. Particularly, since teenagers are in a critical developmental stage where they are learning how to socialize and developing an identity [52], this kind of design could assist with how they navigate socialization when gaming with others. Discord and gaming platforms could help players make more informed decisions about collaborating with unfamiliar players in-game. This would support cross-platform practices by enabling players to signal interests such as trading availability and communication preferences or use badges indicating whether they prioritize social interaction or gameplay focus.

Designs for family involvement may be particularly applicable for teenage users. This could prevent risky behavior and to encourage healthier online interactions [55]. Rather than relying on parental monitoring of kids' online activities [81], family involvement through co-playing games could create a meaningful family experience [99]. This shared experience of cross-platform play simultaneously functions as an attentional safeguard for teenagers because it cultivates "infrastructural competence" by using strategies that seamlessly integrate multiple platforms [56].

Tools could actively capture players' attention during high-risk moments, such as trust trades, when decisions could compromise account security or player safety. Prior research suggests that interventions for user' privacy and security must go beyond improving awareness to designing tools that support attention and habit formation during high-risk moments [4, 5]. Similarly, our study shows that security vulnerabilities stem not only from gaps in player awareness, but from the interplay between increasingly sophisticated attacks and the cognitive demands of gaming. For instance, to prevent phishing attacks that compromised P6 and P12's game and Discord accounts, each platform could check participants' attention and notify them that trades should not require the player to log into any additional platform.

Bridging the platform gap would not only mitigate cross-platform risks but also strengthen trust in each platform, benefit the gaming industry, and foster healthier social interactions for players. Cross-platform design is critical because platform-to-platform interactions and their design implications shape users' ability to develop social connections with one another [141]. By proactively addressing the platform gap between in-game spaces and external platforms, like Discord, multiple stakeholders can foster safer, more transparent, and more socially supportive gaming ecosystems.

## 7 LIMITATIONS AND FUTURE WORK

Our findings are based on interviews with 16 teenage Discord users and therefore do not represent the full diversity of Discord's global user base, which spans millions of users across age groups and communities [82]. As a qualitative study centered on teenagers' cross-platform gaming and social experiences, our analysis is limited in its ability to generalize to adult users or to broader populations with different gaming practices. Additionally, our sample reflects limited demographic diversity, which may influence the types of cross-platform interactions and risks observed. While participants self-identified their race, we intentionally avoid making broad cultural assumptions based on these demographics alone. Instead, we situate our findings directly within the participants' lived gaming experiences as shared in their narratives. We also recognize the ethical and methodological constraints of studying youth experiences in environments where social interaction, identity disclosure, and exposure to scams are ongoing risks, which may shape how participants interpret and report their experiences. We acknowledge that peer group norms among teenagers are highly variable and context-dependent; however, their cross-platform practices identified may still offer transferability to similar online gaming communities and youth spaces.

We focus on teenagers' Discord-based gaming practices rather than their interactions with platform moderation systems. This choice may limit our ability to fully capture how governance mechanisms operate across other gaming platforms. Future work should examine a wider range of ages, backgrounds, and gaming communities to better understand variation in cross-platform practices and risks. It could also further investigate how moderation and safety systems function within what we describe as a gaming platform gap, particularly in relation to player needs and cross-platform harm.

## 8 CONCLUSION

Through semi-structured interviews with teenage users of Discord, our research highlights how Discord impacts social gaming experiences. Their experiences demonstrate the prevalence of Discord in gaming activities, such as enhancing competitive cooperation and providing timely game announcements. Moreover, their cross-platform practices supported social interaction, where participants maintained online and offline relationships, organized gameplay, and engaged in social support. Our study of teenagers' social gaming experiences on Discord provides new insight into the platform gap. Gamers could be harmed when engaging in cross-platform practices, like being scammed or hacked. As Discord continues to grow in popularity, the multi-platform gaming ecosystem must strengthen its oversight of emerging forms of cross-platform harm. We therefore urge Discord and game developers to take proactive steps to prevent player abuse by identifying, monitoring, and moderating harmful activities that span multiple platforms and occur within their services.

## ACKNOWLEDGMENTS

We sincerely thank all the teenagers who participated in this study for openly sharing their experiences with risky social interactions on Discord. We are also grateful to their parents for taking the time to understand the study and provide consent for their teenagers' participation. Finally, we thank the anonymous reviewers for their thoughtful feedback and valuable suggestions, which helped improve this paper.

This material is based upon work supported by the U.S. National Science Foundation under award No. 2334934. Any opinions, findings and conclusions or recommendations expressed in this material are those of the author(s) and do not necessarily reflect the views of the U.S. National Science Foundation.

## A1 DEMOGRAPHICS

Table 2: Participant Demographics

| ID | Age | Gender | Racial or Ethnic Identity[a] | Neighborhood Type | Games Played |
|---|---|---|---|---|---|
| P1 | 17 | Female | Asian/Korean | Urban | Valorant, League of Legends |
| P2 | 14 | Male | Asian | Suburban | League of Legends |
| P3 | 15 | Female | Asian | Urban | N/A |
| P4 | 15 | Female | Asian | Suburban | Roblox, Animal Crossing |
| P5 | 14 | Male | White | Suburban | Fortnite, Steam Games (variety) |
| P6 | 15 | Female | Asian | Urban | Roblox |
| P7 | 16 | Female | White | Suburban | Minecraft |
| P8 | 13 | Female | White | Rural | Roblox |
| P9 | 15 | Male | Asian | Urban | Valorant, Rhythm Games |
| P10 | 14 | Male | Asian | Suburban | Roblox, Valorant |
| P11 | 14 | Female | Asian | Urban | Roblox |
| P12 | 16 | Male | Asian | Suburban | Minecraft |
| P13 | 17 | Male | Asian | Suburban | Roblox, Minecraft |
| P14 | 14 | Non-binary | White | Suburban | Roblox |
| P15 | 17 | Male | White | Suburban | Minecraft, Rust, Clash Royale, Valorant |
| P16 | 14 | Female | Mexican/ European | Suburban | Roblox |

[a] Participants were encouraged to self-describe their background along racial or ethnic lines that they preferred.

## A2 INTERVIEW PROTOCOL

**Experiences using Discord:**

1. How long have you been using Discord and what made you start using it?
2. What do you usually do when talking with new people on Discord?
3. Can you share an experience on Discord that you found difficult or unexpected?

**Risky Social Interactions on Discord:**

4. Can you describe any experiences on Discord that you found concerning or uncomfortable?
    a. How do you deal with this?
5. Have there been any upsetting feelings over something that happened on Discord?
6. Is there a time when you had an uncomfortable experience with another user?
7. If a connection on Discord isn't working out, how do you usually handle it?
8. Can you describe a time you met someone on Discord who was difficult to deal with?

**Discord Safety Design Questions:**

9. Can you share what makes you feel safe in servers? Do you use any tools or settings?
10. If you could change or add anything to how safety works on Discord, what would that be?
11. Can you describe anything you learned from using Discord's safety tools?